\documentclass[a4paper,12pt]{article}
\usepackage[dvips]{graphicx}
\DeclareGraphicsExtensions{.pdf,.jpg}

\usepackage{amsfonts,amsmath,amsthm,amssymb}
\usepackage{mathrsfs}
\usepackage[onehalfspacing]{setspace}
\author{L. Ort\'{i}z\footnote{leonardo.ortiz@uniofyorkspace.net}}

\title{A note on the two point function on the boundary of AdS spacetime}

\begin{document}

\maketitle

\begin{center}

Department of Physics\\
University of Guanajuato\\
Leon Guanajuato 37150, Mexico\\\vspace{0.4cm}

\normalsize{\textbf{Abstract}}\\\end{center} \small{We calculate by a new way the two point function on the boundary of AdS spacetime in 1+2 dimensions for the massless conformal real scalar field. The result agrees with the answer provided by the Boundary-limit Holography and Witten recipe. This is done in Poincar\'{e} coordinates. The basic ingredients of this new method are conformal techniques, quantum fields defined on a half of Minkowski spacetime and a limit inspired by the Boundary-limit Holography. We also show that a state in AdS, the global vacuum, in three dimensions induces a state on the conformal boundary of AdS spacetime, which in turn induces a state on the BTZ black hole. On the other hand the same state in AdS induces a state on the BTZ black hole which in turn induces a state on its conformal boundary. The two ways of getting the state on the conformal boundary of the BTZ black hole coincide for the massless conformal real scalar field. We point out that the normalizable modes in the AdS/CFT correspondence for the BTZ black hole give a similar contribution as the non-normalizable modes used in the Witten prescription. We also give some clues on why the Witten and the Boundary-limit Holography prescription coincide.}\vspace{0.3cm}

%PACS numbers: 04.62.+v, 03.70.+k\newpage

\newpage

\section{Introduction}

The Maldacena conjecture \cite{jMald98}, without doubt, has been a breakthrough in modern theoretical physics. It has set up a new paradigm in the way how we consider theories living in AdS spacetime. Shortly after this conjecture, also known as AdS/CFT correspondence, was posed, Witten \cite{eWitt98} and Gubser \textit{et al.} \cite{sGub98} put it in field theoretical terms. Inspired by these works, several other approaches to the AdS/CFT correspondence have been arisen, see for example \cite{gKeri06} and references therein. In particular, Algebraic Holography \cite{khReh00} and the Boundary-limit Holography \cite{mBerjBrosuMosrSch00} are examples where the background in fixed and one, effectively, is studying QFT in AdS spacetime and relating it to conformal QFT in its conformal boundary. In this context, QFT in curved spacetime techniques are relevant and arise naturally in the AdS/CFT correspondence. Also some past works on QFT in AdS spacetime \cite{sjAvjIshdSto78}, \cite{BreiFreed82}, \cite{BreiFreed822} become relevant \cite{KleWitt99} for the Maldacena conjecture. It is fair to say that the approach were one uses QFT in curved spacetime techniques in the correspondence is not in the mainstream however there are important works and results that have been obtained within this approach. We feel that it is good to make a bridge between the different approaches to the AdS/CFT. One purpose of this work is to contribute to make this bridge, in particular we will give several clues on how the Witten approach and the Boundary-limit Holography are related.

One of the objects one would like naturally to calculate in the CFT side of the AdS/CFT correspondence are the n-point functions and a recipe for this was given in \cite{eWitt98}. This method involves the partition function in the bulk and in the boundary. Inspired in part by this work, Bertola \textit{et al.} \cite{mBerjBrosuMosrSch00} proposed a method for obtaining n-point functions of a CFT theory living in the conformal boundary of AdS spacetime from n-point functions of the real scalar field living in AdS spacetime. Both methods seem to be different but give the same answer, at least for the real scalar field living in AdS spacetime. The reason for this coincidence has not been clarified so far although there has been studies in this direction \cite{mDuekhReh02}. In this work we propose and different way of getting the two point function for the massless conformal real scalar field. In this method we use techniques from QFT in curved spacetime. We do the calculation for the two point function and in three dimensions but we do not see any obstacle, apart from technicalities, to obtain any n-point function in any dimension.

In the mainstream AdS/CFT literature it has been proposed \cite{vBala98}, \cite{uhDan99} that at the level of QFT states in the bulk correspond to states in the boundary in the AdS/CFT correspondence, and Algebraic Holography \cite{khReh00} and the Boundary-limit Holography \cite{mBerjBrosuMosrSch00} confirm this. In this context there is a map between states in the bulk and states in the boundary. Now, in field theory besides vacuum (ground) states thermal states are outstanding in the theory. In the AdS/CFT correspondence they arise naturally when we have a black hole living inside the bulk. In the three dimensional case they have arisen in the work of Keski-Vakkuri \cite{eKe99} and \cite{lOrt13}. In the first of these works it was used the Witten prescription for computing the state on the boundary from the theory in the bulk. In this calculation Keski-Vakkuri used only the contribution of the non-normalizable modes and discarded the normalizable modes. In the present work we will show that the normalizable modes, under the prescription of the Boundary-limit Holography, give a similar contribution to the theory in the boundary of the BTZ black hole for the massless conformal real scalar field. This fact has been done already in \cite{lOrt13} by a different method.

At the end of the present work, based on \cite{mSpraaStro99}, we give an analysis of the way one gets the two point function in the Witten prescription and compare it with the Boundary-limit Holography prescription \cite{mBerjBrosuMosrSch00}. This sheds some light in why the two methods coincide at least for the real scalar field. We consider this analysis is important since it could help to put the AdS/CFT correspondence in a more rigourous setting.

The organization of this work is as follows: in section 2 we present a new way of getting the two point function on the conformal boundary of AdS spacetime by using techniques of QFT in curved spacetime and for the massless conformal real scalar field. In section 3 we show how to obtain a thermal state on the boundary of the BTZ black hole by using the Boundary-limit Holography and past work of QFT in BTZ black holes. In section 4 we analize the procedure for obtaining the two point function in the Witten prescription and compare it with the Boundary-limit prescription. Finally, in section 5 we give some conclusions of this work.

\section{A new way to get the two point function }

Clearly the two point function is a very important object in any QFT. In this section we get the two point function on the boundary of AdS spacetime in three dimensions for a massless conformal real scalar field in the bulk. In doing this we use conformal techniques and ideas from QFT in curved spacetime.

Witten and Bertola \textit{et al.} use Poincar\'{e} coordinates in their works. In these coordinates, in 1+2 dimensions, the line element of AdS spacetime is
\begin{equation}\label{E:1}
ds^{2}=\frac{\ell^2}{z^2}\left(-dT^{2}+dk^{2}+dz^{2}\right),
\end{equation}
where $z\in(0,\infty)$, $k\in(-\infty,\infty)$, $T\in(-\infty,\infty)$ and $\ell$ is related with the cosmological constant as $\Lambda=-\frac{1}{\ell^2}$. By looking at this line element is evident that the Poincar\'{e} chart is conformal to a half of the Minkowski chart. Inspired by this and by works \cite{bDW75}, \cite{ndBpcwD82} where there appears quantum fields defined on a half of the Minkowski spacetime, one naturally asks if it is not possible to apply conformal techniques and fields defined just on a half of the Minkowski spacetime to obtain a two point function on the conformal boundary of AdS spacetime in 1+2 dimensions, from the two point function associated with the conformal vacuum of the real scalar field obtained from the vacuum defined on a half of spacetime which is conformal to the Poincar\'{e} chart. The answer turns out to be positive and coincides with
\begin{equation}\label{E:140ll}
F_{b}(\Delta T,\Delta k)=\frac{1}{4\ell\pi}\lim_{\epsilon_{+}\rightarrow 0}\frac{1}{((\Delta
k)^{2}-(\Delta T+i\epsilon)^{2})^{\frac{3}{2}}},
\end{equation}
which is the result from Boundary-limit Holography and Witten recipe, where $\Delta T=T-T'$ and $\Delta k=k-k'$. The purpose of this section is to report this result. The equation (\ref{E:140ll}) is the two point function obtained \cite{lOrt13} by using Boundary-limit Holography \cite{mBerjBrosuMosrSch00} for the massless conformal scalar field. This result is interesting and relevant for the AdS/CFT literature since the two point function in the AdS/CFT duality is obviously a very important object in the theory, and in getting it we are just using techniques from QFT in curved spacetime and a limit inspired by Boundary-limit Holography. Also we mention that in principle the same method could be applied to obtain any n-point function.

Now, by a new method, we get (\ref{E:140ll}) by
starting with a set of normalized solutions of the Klein-Gordon
operator with metric
\begin{equation}\label{E:106}
ds^{2}=-dT^{2}+dk^{2}+dz^{2},
\end{equation}
where the coordinates have the same range as Poincar\'{e}
coordinates. One set of such solutions is
\begin{equation}\label{E:322}
F_{\omega na}'(T,k,z)=\frac{1}{\pi(2\omega)^{1/2}} e^{-i\omega
T}e^{i nk}\sin(az),
\end{equation}
where $\omega=\left(n^{2}+a^{2}\right)^{1/2}$, $-\infty<n<\infty$
and $a>0$. By choosing the coupling constant $\xi=\frac{1}{8}$,
then a set of solutions of the Klein-Gordon operator on AdS
spacetime in Poincar\'{e} coordinates is
\begin{equation}\label{E:323}
F_{\omega na}(T,k,z)=\frac{z^{1/2}}{\ell^{1/2}}\frac{1}{\pi(2\omega)^{1/2}}
e^{-i\omega T}e^{i nk}\sin(az).
\end{equation}
Then a real linear scalar field on AdS spacetime can be expanded
as
\begin{equation}\label{E:324}
\hat{\phi}(T,k,z)=\int_{-\infty}^{\infty}\int_{0}^{\infty}\left(F_{na}\hat{a}_{na}+F^{*}_{na}\hat{a}^{\dagger}_{na}\right)dnda,
\end{equation}
where the conformal vacuum is defined by
\begin{equation}\label{E:325}
\hat{a}_{na}|0\rangle=0\hspace{0.5cm}\forall\quad n, a.
\end{equation}
From (\ref{E:324}) and (\ref{E:325}) it follows that the two point
function is given by
\begin{eqnarray}\label{E:328}
\langle 0|\hat{\phi}(T,k,z)\hat{\phi}(T',k',z')|0\rangle
&=&\frac{(zz')^{1/2}}{8\ell\pi}\times\nonumber\\
&\times &\lim_{\epsilon_{+}\rightarrow 0}
\left(\frac{1}{\left((\Delta k)^{2}-(\Delta
T+i\epsilon)^{2}+(z-z')^{2}\right)^{1/2}}-\right .\nonumber\\
&-&\left .\frac{1}{\left((\Delta k)^{2}-(\Delta
T+i\epsilon)^{2}+(z+z')^{2}\right)^{1/2}}\right),
\end{eqnarray}
where $\Delta T=T-T'$ and $\Delta k=k-k'$ and we have used that
$2\sin(az)\sin(az')=\cos(a(z-z'))-\cos(a(z+z'))$ and the integrals 3.961 2 and
6.671 6 in \cite{asGradimRyz80}. Now, if we expand the last expression for small $z$ and $z'$ and take the limit when they go to zero and at the same time multiply all the expression for $(zz')^{-{\frac{3}{2}}}$ we finally get
\begin{equation}\label{E:2}
\langle 0|\hat{\phi}(T,k)\hat{\phi}(T',k')|0\rangle=\frac{1}{4\ell\pi}\lim_{\epsilon_{+}\rightarrow 0}\frac{1}{((\Delta
k)^{2}-(\Delta T+i\epsilon)^{2})^\frac{3}{2}}.
\end{equation}

By comparing this equation and (\ref{E:140ll}) we see that both equations are the same, so Boundary-limit Holography \cite{mBerjBrosuMosrSch00}, Witten recipe \cite{eWitt98} and our method give the same answer. It is important to note that the power of the factor $zz'$ by which we multiply the two point function is the same in both cases, so in the method presented in this section somehow we are using the same limit as Boundary-limit Holography.

Since the Boundary-limit Holography and our method use as a basic element the geometry of AdS spacetime we can say that it fixes the form of the two point function for the real scalar field. We point out that, in principle, this method can be extended to obtain any n-point function and can be applied to the AdS/CFT correspondence in the semiclassical limit. Also the result presented here shows that quantum theory in curved spacetime is relevant for semiclassical issues of the AdS/CFT correspondence. It would be interesting to extend this work to other dimensions and to other fields such as the Dirac or the electromagnetic field.

\section{The boundary limit of QFT on the BTZ black hole}

On general grounds, in the context of the AdS/CFT correspondence, one expects that a black hole sitting inside AdS should be dual to a thermal state on the conformal boundary of the black hole. This has indeed been obtained for the BTZ black hole by Keski-Vakkuri \cite{eKe99}. In this work the Witten prescription \cite{eWitt98} to get the state on the boundary was used and the normalizable modes were not taken into account. However from the analysis of Balasubramanian \textit{et al.} \cite{vBala98} one knows that the normalizable modes are indeed important in the AdS/CFT correspondence, since they are dual to states on the conformal boundary. In this section we show that the normalizable modes contribute with a similar amount to the state on the boundary under the prescription of the Boundary-limit Holography. This in particular shows a relation of past work on QFT in BTZ black holes with the AdS/CFT correspondence. As far as we know no one else before has pointed out this relation. The fact that the normalizable modes contribute on equal footing to the state on the boundary as the non-normalizable modes has been implicitly done in \cite{lOrt13}. However in \cite{lOrt13} we did QFT in AdS then went to the boundary by using the Boundary-limit Holography \cite{mBerjBrosuMosrSch00} then took the restriction to the exterior of the BTZ black hole. In the present section we follow a different procedure, we have QFT in AdS then QFT in the BTZ black hole and the go to the boundary under the prescription of the Boundary-limit Holography \cite{mBerjBrosuMosrSch00} applied to the BTZ black hole. This suggests that, in certain cases, the following two procedures are equivalent:
\begin{equation}\nonumber
\textrm{QFT in bulk AdS}\rightarrow \textrm{Boundary-limit} \rightarrow \textrm{Restriction to the BTZ black hole}
\end{equation}
\begin{equation}\nonumber
\textrm{QFT in bulk AdS}\rightarrow \textrm{Restriction to the BTZ black hole}\rightarrow \textrm{Boundary-limit}
\end{equation}

In both procedures we get a thermal state given by
\begin{equation}\label{E:1}
F_{b}(\Delta
t,\Delta\phi)\sim\sum_{n\in\mathbb{Z}}\frac{1}{(\cosh\kappa\ell(\Delta\phi+2\pi
n)-\cosh\kappa(\Delta t-i\epsilon))^{\frac{1}{2}}},
\end{equation}
where $\kappa$ is the surface gravity, $\ell$ is related with the cosmological constant as $\Lambda=-\frac{1}{\ell^2}$ and $\Delta t=t_{1}-t_{2}$, $\Delta \phi=\phi_{1}-\phi_{2}$ are the coordinates inherited from the spacial BTZ black hole coordinates.

This is telling us that a state on AdS, the global vacuum, induces a state on its conformal boundary which in turn induces a state on the BTZ black hole; and this is equivalent to have the same state on AdS spacetime then a state on the BTZ black hole then a state on the conformal boundary of the BTZ black hole

Now we show that we get (\ref{E:1}) by two distinct procedures for the massless conformal real scalar field. This is done by taking into account our previous work \cite{lOrt13} and the work of Lifschytz-Ortiz \cite{gLifmOrt94}.

From (61) in \cite{lOrt13} we see that if $p=-\frac{1}{2}$ then we get (\ref{E:1}). In this case (61) in \cite{lOrt13} was calculated with the following procedure: QFT in bulk AdS, in Poincar\'{e} coordinates, then the Boundary-limit \cite{mBerjBrosuMosrSch00}, and then restriction to the exterior of the BTZ black hole in the boundary. We see that $p=-\frac{1}{2}$ corresponds to take the massless conformal scalar field. In this case we are assuming the global vacuum is the same that the Poincar\'{e} vacuum. This is a valid assumption since it has been shown \cite{uhDan99} that for most purposes they can be considered equivalent.

Now let us discuss the other procedure. According to \cite{gLifmOrt94} the two point function for the massless conformal real scalar field on the exterior of the BTZ black hole is given by
\begin{equation}\nonumber
G(x,x')=\sum_{n\in\mathbb{Z}}\frac{1}{\left(\frac{rr'}{r_{+}^2}\cosh\kappa\ell(\Delta\phi+2\pi
n)-1-\frac{(r^2-r_{+}^2)^\frac{1}{2}(r'^2-r_{+}^2)^{\frac{1}{2}}}{r_{+}^2}\cosh\kappa(\Delta t-i\epsilon)\right)^{\frac{1}{2}}}.
\end{equation}
Now we take the boundary limit of this expression. In order to do this we multiply by $\frac{(rr')^{\frac{1}{2}}}{r_{+}}$ and take the limit when $r,r'\rightarrow \infty$. In this limit we get (\ref{E:1}) again. In this case we have: QFT in bulk AdS spacetime, then restriction to the exterior of the BTZ black hole and then Boundary-limit. So we have two procedures to get the same state. These two procedures use only QFT in curved spacetime techniques and the Boundary-limit Holography ideas \cite{mBerjBrosuMosrSch00}.

\section{Are Witten and the Boundary-limit Holography prescriptions equivalent?}

 On one hand the Witten prescription \cite{eWitt98} for getting n-point functions in the CFT side of the AdS/CFT correspondence uses the bulk-boundary propagator. On the other hand the Boundary-limi Holography prescription \cite{mBerjBrosuMosrSch00} for getting n-point functions in the boundary uses the n-point functions in the bulk and a limit process in passing from the bulk to the boundary. A limit process, which would require regularization, is also involved in the Witten prescription. However when one sees both methods at work it is not obvious that they are related or how they are related. In this section we will give some clues on how they are related. We will base our analysis on \cite{mSpraaStro99}.

 For the convenience of the reader we reproduce here some equations, the more relevant for our purposes, from \cite{mSpraaStro99}. The Witten prescription essentially entail the following steps: take the two point function in the bulk which you calculate using QFT in curved spacetime. We denote this two point function as $G(z, t, x;z', t', x')$. Here we are taking Poincar\'{e} coordinates $(z, t, x)$ and $x$ denotes the spacial coordinates of the Minkowski spacetime in the boundary of AdS spacetime, however the same analysis can be done in other coordinates although the Poincar\'{e} chart is the most natural for AdS/CFT correspondence purposes. We also have in mind the real scalar field since the Boundary-limit Holography has been tested in this case. Now define the bulk-boundary propagator as
 \begin{equation}\label{E:11}
 K(z, t, x;t', x')=\lim_{z'\rightarrow 0}[(z')^{-h}G(z, t, x;z', t', x')],
 \end{equation}
 where $h$ is solution of $m^{2}=h(h-1)$ and $m$ is the mass of the field. It is clear here that the choice of vacuum is very relevant for the Witten prescription. It is also, obviously, important for the Boundary-limit Holography prescription. Now if we have some data $\phi_{0}(t,x)$ in the boundary we can extend it to the bulk by
 \begin{equation}
 \phi(z, t, x)=\int dt'dx'K(z, t, x;t', x')\phi_{0}(t', x').
 \end{equation}
 Now we plug this field into the usual action of the real scalar field. Then after using the field equations we get
 \begin{equation}\label{E:13}
 S=\lim_{z\rightarrow 0}\left[\frac{1}{2}\int dtdx\phi(z, t, x)\partial_{z}\phi(z, t, x)\right].
 \end{equation}
 Now it is assumed that the bulk-boundary propagator approaches a delta function as $z\rightarrow 0$
 \begin{equation}
 K(z, t, x;t', x')\rightarrow z^{-h+1}\delta(t-t')\delta(x-x').
 \end{equation}
 Hence
 \begin{equation}
 \phi(z, t, x)\rightarrow z^{-h+1}\phi_{0}(t,x).
 \end{equation}
 In this conditions equation (\ref{E:13}) reduces to
 \begin{equation}\label{E:17}
 S=\frac{1}{2}\int dtdxdt'dx'\phi_{0}(t,x)\phi_{0}(t', x')[\lim_{z\rightarrow 0}z^{-h+1}\partial_{z}K(z, t, x;t', x')].
 \end{equation}
 Then using the partition function one finally gets
 \begin{equation}\label{E:18}
 \langle \mathcal{O}(t,x)\mathcal{O}(t', x')\rangle=\lim_{z, z'\rightarrow 0}[(z')^{-h}z^{-h+1}\partial_{z}G(z, t, x;z', t', x')].
 \end{equation}
 This equation is already in (\ref{E:17}) inside the square brackets. So from this point of view taking the functional derivative in the Witten prescription has the effect of just taking out the two point function, it is just a trick. However as we assume \cite{mSpraaStro99} that $G(z, t, x;z', t', x')$ goes like $z^{h}$ as either of the two points go to the boundary then the effect of the derivative in (\ref{E:18}) is just to get a factor $z^{h-1}$. So it would have the same effect if we just multiply $G(z, t, x;z', t', x')$ by $(z')^{-h}z^{-h}$ and take the limit $z, z'\rightarrow 0$. But this is precisely the Boundary-limit Holography prescription \cite{mBerjBrosuMosrSch00}. Hence this analysis explains why both prescriptions give the same answer. However the Boundary-limit Holography is valid for interacting fields too.

\section{Conclusions}

The calculations and arguments presented in this work confirm that QFT in curved spacetime is relevant for the AdS/CFT correspondence. In works such as Algebraic Holography \cite{khReh00} and the Boundary-limit Holography \cite{mBerjBrosuMosrSch00} it is clear from the outset that it was so. However with the results present in this work we emphasize that there is a relation between QFT theories in AdS spacetime and CFT in its conformal boundary. More particularly, the calculation of section 2 shows that the geometry of AdS spacetime fixes the two point function for the massless conformal real scalar field. It would be interesting to generalize this to other fields.

In section 3 we related old work on QFT in the BTZ black hole with work on the AdS/CFT correspondence and we pointed out the fact that the normalizable modes contribute in a similar way as the non-normalizable modes. We think this is important since Keski-Vakkuri \cite{eKe99} in his calculation neglected the normalizable modes. With the present work is clear that you can not throw away the normalizable modes, this reflect the spirit which Balasubramanian \textit{et al.} have \cite{vBala98}. Now if we interpret that the restriction of the global vacuum, which corresponds to the Hartle-Hawking-Israel vacuum of the BTZ black hole \cite{gLifmOrt94}, to the exterior of the BTZ black hole as a thermalization of the field, then the same happens in the boundary. So this is telling us that in some sense a kind of Unruh effect is happening in the bulk and in the boundary but in the bulk it has the interpretation of the thermal state of the black hole; whereas in the boundary it can be given the normal interpretation of a thermal state for an accelerated observer.

It has been pointed out \cite{mDuekhReh02} that the Boundary-limit Holography \cite{mBerjBrosuMosrSch00} and the Witten prescription \cite{eWitt98} give the same result and the reason for this coincidence has not been clarified, although some analysis in this direction have been carried out in \cite{mDuekhReh02}. In this work we have shown more clearly the reason of this coincidence. So section 4 can be considered as complementary to \cite{mDuekhReh02}. For practical purposes we can just restrict to take boundary limits of n-point functions in the bulk and take them as n-point functions in the boundary. It seems that the introduction of the partition function in Witten recipe has a physical motivation but at the end it is a kind of trick to take out the n-point functions, it is in the same spirit as the introduction of a current in QFT for obtaining the expectation value of time order product of operators. Now if we take the present work together with the Boundary-limit Holography \cite{mBerjBrosuMosrSch00} we can say that the Witten prescription has been put in a more rigourous setting. We would like to finish by saying that the Witten prescription has already half of the Boundary-limit Holography in the definition of the bulk-boundary propagator (\ref{E:11}), the other half is introduced in taking the limit of the action.

\section*{Acknowledgments}

We thank A. Higuchi and B. S. Kay for their comments on section 2 of this work and for
their help in passing from (\ref{E:328}) to (\ref{E:2}). We also thank Prof. O. Obreg\'{o}n for his support.\vspace{0.3cm}

This work was sponsored by CONACYT-Mexico through a postdoctoral fellowship.

\end{document}